\documentclass[aps,reprint,notitlepage,onecolumn,floatfix,nofootinbib]{revtex4-1}

\newcommand{\vk}{{\bf k}}

\begin{document}
\title{Signatures of Brane Inflation}
\author{Ashok Thillaisundaram}
\email{A.Thillaisundaram@damtp.cam.ac.uk}
\affiliation{Department of Applied Mathematics and Theoretical Physics, University of Cambridge,\\
Cambridge CB3 0WA, England}

\begin{abstract}
We study slow-roll inflation on a three-brane in a five-dimensional bulk where the effects of energy loss from the brane due to graviton emission is included in a self-consistent manner. We explicitly derive the form of the energy loss term due to inflaton-to-graviton scattering and thus determine the precise dynamics of the two resulting inflationary solutions. What is also remarkable is that nonconservation of energy on the brane causes the curvature perturbation to not be conserved on superhorizon scales even for the purely adiabatic perturbations produced in single-field inflation. Thus the standard method of calculating the power spectrum of inflaton fluctuations at Hubble exit and equating it to the power spectrum at horizon reentry no longer holds. The superhorizon evolution of the perturbations must be tracked from horizon exit through to when the modes reenter the horizon for the late time power spectrum to be calculated. We develop the methodology to do this in this paper as well.
\end{abstract}

\maketitle
\nopagebreak

\section{Introduction}

Recent progress in superstring theory and  M-Theory has led to increased interest in cosmological models with extra dimensions. Braneworld models are a particular class of these models in which our four-dimensional Universe lies on a three-brane embedded in a higher-dimensional (possibly large) spacetime (referred to as the bulk) \cite{Davis2004}. Matter fields are confined to the brane while gravity is free to propagate in the bulk. An example of these models which has been very well studied is the one proposed by Randall and Sundrum where a $Z_2$-symmetric (i.e. mirror symmetric) three-brane is embedded in five-dimensional Anti-de Sitter (AdS) spacetime \cite{RandallSundrum1999b}.

However, as mentioned, gravity is not confined to the brane, and so fluctuations in the matter fields on the brane can emit gravitational waves (or bulk gravitons) into the bulk. The brane energy-momentum tensor is therefore not conserved and there will be a flow of energy from the brane to the bulk. It is essential that this effect be considered for a cosmological model to be realistic. And in the context of the Randall-Sundrum II model, it has been realised that the bulk cannot simply be pure AdS but must also have a bulk energy-momentum tensor describing the radiation of gravitons from the brane. This has been studied by several authors. For example in \cite{HebeckerMarch-Russell2001}, the authors consider thermal emission of gravitons from the brane. In \cite{LangloisSorboRodriguez2002,LangloisSorbo2003} on the other hand, an exact bulk solution is used, the five-dimensional generalisation of Vaidya's metric, which describes gravitons being emitted orthogonally from the brane.

These two effects, namely the presence of extra dimensions and the nonconservation of energy-momentum on the brane, may significantly affect our picture of the early Universe. As such, inflation (and the spectrum of perturbations generated) offers a powerful method of testing such a model. The spectrum of inflationary perturbations is likely to be modified due to these two effects. Common observables like the spectral index for this model can therefore be calculated and confronted with astronomical observations.

Inflation on a brane within a higher dimensional spacetime has already been extensively studied \cite{MaartensWands2000,Rubakov2007}. However, to the best of our knowledge, the effect of the emission of gravitons from the brane (i.e. the nonconservation of brane energy-momentum) on inflation has only recently been considered \cite{BraxDavis}. The purpose of this paper is therefore to build on the work done in \cite{BraxDavis} and to further investigate novel consequences of nonconservation of brane energy-momentum on a canonical model of slow-roll chaotic inflation on the brane. It is assumed that the inflaton is confined to the brane and that it is the dominant source of energy-momentum. The effect on the brane-bulk system of graviton emission from the brane is included in a self-consistent manner, as was done in \cite{LangloisSorboRodriguez2002}.

 This paper is organised as follows. In section 2, we present the governing equations of the model. The slow-roll inflationary background equations are then the focus of section 3 and the precise dynamics of the two resulting inflationary solutions are derived. In section 4, we consider scalar perturbations about the homogeneous background. In particular, we demonstrate that the curvature perturbation is not conserved on superhorizon scales even for adiabatic perturbations and that new tools must be developed to calculate the late time power spectrum. We end with our conclusions in section 5. 

\section{Basic Model}
\subsection{Governing Equations}

We follow the model presented in \cite{LangloisSorboRodriguez2002}. This model consists of a $Z_2$-symmetric three-brane, with tension $\lambda$, embedded in a five-dimensional bulk spacetime. Ordinary cosmological matter is present on the brane with energy density $\rho$ and pressure $p$ (later, when we consider inflation, the inflaton field will be the dominant component on the brane). The five-dimensional bulk spacetime has negative cosmological constant $\Lambda=-6\mu^2$ and an energy-momentum tensor given by:
\begin{equation}
T_{AB}=\sigma k_A k_B,
\end{equation}
where $k^A$ is a null vector orthogonal to the brane and $\sigma$ represents the energy flux of the bulk gravitons from the perspective of a brane observer. We assume that the five-dimensional Einstein equations hold in the bulk, thus the bulk spacetime satisfies
\begin{equation}
G_{AB}+\Lambda g_{AB}=\kappa_5^2 T_{AB}.
\end{equation}
Here $\kappa_5^2\equiv M_5^{-3}$ is the five-dimensional gravitational coupling.
\\
\\
These five-dimensional Einstein equations can be solved using the following metric ansatz
\begin{equation}
ds^2=-f(r,v)dv^2+2dr dv +r^2 d\textbf{x}^2,\text{     } f(r,v)=\mu^2 r^2 - \frac{\mathcal{C}(v)}{r^2}.
\end{equation}
This is the five-dimensional generalisation of Vaidya's metric \cite{Vaidya} which describes the spherically-symmetric spacetime surrounding a radiating star. The coordinate $r$ is identified with the scale factor $a$ on the brane. It should be noted that the price we have to pay for having an exact analytic solution for the bulk spacetime is that we must assume that all the gravitons are emitted ``radially'' from the brane, an assumption which is not likely to be true at high energies. (Here, the radial direction $r$ essentially means the direction orthogonal to the brane in the bulk spacetime.)

Using various components of the five dimensional Einstein equations, as well as the Israel junction conditions at the brane location, we obtain the following equations:

\begin{equation}\label{Equation:ModifiedFriedmann}
H^2=\frac{\kappa_5^4}{36}\rho^2 + \frac{\kappa_5^4}{18}\lambda\rho + \frac{\mathcal{C}(v)}{a^4}
\end{equation}
\begin{equation}
\frac{d\mathcal{C}}{dv}=\frac{2\kappa_5^2 \sigma}{3}a^3\big( \dot{a}-\sqrt{f+\dot{a}^2}\big) ^2
\end{equation}
\begin{equation} \label{Equation:Nonconservation}
\dot{\rho}+3H(\rho+p)=-2\sigma.
\footnote{The vector $k^A$ and the unit normal $n^A$ which would appear in the junction conditions have been selected and normalised appropriately so that these equations appear in the above nice form. More details on this are given in \cite{LangloisSorboRodriguez2002}.}
\end{equation}
Before we proceed, some comments should be made regarding this system of equations. First, for the modified Friedmann equation (\ref{Equation:ModifiedFriedmann}), if we use the Randall-Sundrum condition ($\kappa_5^2\lambda=6\mu$) and make the following identification,
\begin{equation}
\label{masses}
\frac{\kappa_5^4 \lambda}{6}=\kappa_5^2\mu\equiv 8\pi G=M_p^{-2},
\end{equation}
then the second term agrees with the standard Friedmann equation. Also, by comparing the magnitudes of the first and second terms, we can define two regimes: the high energy regime $(\rho\gg  \lambda)$ where $H^2\propto \rho^2$ and the low energy regime  $(\rho\ll  \lambda)$ where $H^2\propto \rho$. The radiation-like term $\frac{\mathcal{C}(v)}{a^4}$ originates from the projection of the bulk Weyl tensor and hence $\mathcal{C}(v)$ is often referred to as the Weyl parameter. If the bulk is empty, then $\mathcal{C}(v)$ is constant and the 5-dimensional metric is equivalent to the AdS-Schwarzschild metric ($\mathcal{C}$ will therefore be a measure of the mass of the bulk black hole). Here, we have also chosen to set the four-dimensional cosmological constant to zero; we do not attempt to solve the cosmological constant problem with this model.

And lastly, for the nonconservation of energy equation (\ref{Equation:Nonconservation}), the term on the RHS represents the loss of energy from the brane into the bulk. The factor $2$ is present because the brane is radiating on both sides, into two copies of the same bulk spacetime.

We are primarily interested in a period of slow-roll inflation during the high energy regime. In this limit, the previous system of equations can be solved. We find that the term $\frac{\mathcal{C}(v)}{a^4}$ is just a very small modification to the Friedmann equation during slow-roll inflation. We therefore choose to neglect it. The system of equations reduces to just

\begin{equation}
H^2=\frac{\kappa_5^4}{36}\rho^2 
\end{equation}
\begin{equation}
\dot{\rho}+3H(\rho+p)=-2\sigma.
\end{equation}

\subsection{Calculation of Loss Term due to Graviton Emission}

We now need to determine the energy loss $\sigma$. We assume that all the energy loss from the brane is due to the process $\phi + \phi \rightarrow G$, i.e. the emission of a bulk graviton from the scattering between two inflaton particles. Our calculation closely follows section III in reference {\cite{LangloisSorboRodriguez2002}. We make several simplifying assumptions. First, it is assumed that the cosmological expansion can be neglected in the computation of the cross-section for this process. A Minkowski background in the brane is therefore assumed. The bulk gravitons can thus be described as (generalised) Kaluza-Klein modes (as done in \cite{LangloisSorboRodriguez2002}). The required amplitude for the process, $\phi + \phi \rightarrow G$, is then given by \cite{LangloisSorboRodriguez2002, Rattazzi}
\begin{equation}
 \vert {\cal M}\vert ^2= \frac{1}{12\pi} \kappa_5^2 s^2
\end{equation}
where the Mandelstam variable is $s= -(E_{k_1}+E_{k_2})^2 + (\vk_1 +\vk_2)^2$. $\vk_{1,2}$ are the momenta of the two incoming inflaton particles and $E_{k_{1,2}}$ represent the energies of the inflaton particles.

To determine the energy loss due to this process, we need to use the relativistic Boltzmann equation with a collision term. However, since we are interested in the energy density on the brane and not the number density of inflaton particles, we must multiply the collision term by the graviton energy and integrate over all Kaluza-Klein modes and over momentum space. The following energy nonconservation equation is obtained:
\begin{equation}
\label{Boltzmann}
 \dot{\rho}+3H (\rho+p)= -\frac{1}{2}\int dm \frac{d^3\bf{k}_{m}}{(2\pi)^3}{\cal  C}
\end{equation}
where 
\begin{equation}
 {\cal C}= \int \frac{d^3\vk_1}{(2\pi)^3 2E_{k_1}}\frac{d^3\vk_2}{(2\pi)^3 2E_{k_2}}\vert {\cal M}\vert^2n_{k_1}n_{k_2} (2\pi)^3 \delta^{(3)} ({\bf{k}_m} -\vk_1 - \vk_2) (2\pi) \delta (E_{\bf{k}_m}-E_{k_1}-E_{k_2}).
\end{equation}
Here, $\bf{k}_{m}$ is the bulk graviton momentum and $n_{k_{1,2}}$ are the number densities of inflaton particles with momenta $\bf{k_{1,2}}$. Similarly, the graviton energy is given by $E_{\bf{k}_m}=\sqrt{\vert{\bf{{k}_m}}\vert^2+m^2}$ where $m$ which labels the Kaluza-Klein mode can be thought of as the graviton mass from the brane perspective. The inflaton particles are treated as effectively massless, so $E_{k_{1,2}}=\vert\bf{k_{1,2}}\vert$.

During inflation we assume that the inflaton fluctuations are in the standard de Sitter vacuum, i.e. the Bunch-Davies vacuum. We further note that for a scalar field in de Sitter space, any timelike observer in the Bunch-Davies vacuum state will measure a thermal distribution of particles with temperature, $T=\frac{H}{2\pi}$ \cite{GibbonsHawking}. We therefore take the number density of inflaton particles to be the thermal distribution:
\begin{equation}
 n_{k_{1,2}}=\frac{1}{e^{\frac{E_{k_{1,2}}}{T}}-1}.
\end{equation}
The RHS of (\ref{Boltzmann}) can then be integrated explicitly\footnote{To obtain this explicit result we must further assume that the contribution to the energy loss is dominated by the ``heavy'' gravitons, i.e. those with $m\gg \mu$. This is explained in reference \cite{LangloisSorboRodriguez2002}.} to give
\begin{equation}
\dot{\rho}+3H (\rho+p)=-\frac{7\cdot5^2 \zeta(\frac{9}{2})\zeta(\frac{7}{2})}{2^{23}\pi^{10}3^8}\kappa_5^{18}\rho^8\equiv-\alpha\kappa_5^{18}\rho^8
\end{equation}
$\alpha$, defined implicitly above, is approximately $3\times10^{-14}$. The $\rho^8$ term is also likely to be very small in magnitude. However, $\kappa_5^{18}$ which is a measure of the difference between the 4D Planck mass, $M_p$ and the 5D fundamental mass scale, $M_5\equiv\kappa_5^{-2/3}$ can potentially be quite large. For example, taking $V=\frac{1}{2}m^2\phi^2 \text{ and } M_5=10^{-3}M_p$, we find that
\begin{equation}
\label{loss}
 \frac{\alpha\kappa_5^{18}\rho^8}{3H (\rho+p)}\approx10^{-4} \left(\frac{M_5}{10^{-3}M_p}\right)^{18\cdot\frac{3}{2}}.
\end{equation}
Thus, the loss term will generally be small, but not necessarily to such a negligible extreme. The dependence of $\kappa_5^{18}$ on the relative sizes of $M_5$ and $M_p$ has been included explicitly in equation (\ref{loss}).

\section{Slow-roll Inflation on the Brane}

\subsection{Two Different Branches of Slow-roll Dynamics}

We consider an inflaton field confined to the three-brane. It is assumed that the inflaton field $\phi$ (with self-interaction potential $V(\phi)$) forms the dominant contribution to the brane energy density and pressure. Our work in this section builds on \cite{BraxDavis}. In \cite{BraxDavis}, the authors discovered that energy loss will result in the presence of a new inflationary branch. Here, we explicitly derive the dynamics of both branches, examine the stability of the new branch, and investigate why a new branch appears when there is energy loss.

We first observe that since the nonconservation of energy equation and the modified Friedmann equation are different to those of standard 4D general relativity, the condition for inflation, i.e. the condition to have accelerating expansion, is now different as well. The following inequality is obtained
\begin{equation}
 \ddot{a}>0\Leftrightarrow p<-\frac{2}{3}\rho\left(1+3\alpha\kappa_5^{16}\rho^6\right),
\end{equation}
which is a stronger condition than the standard 4D result $p<-\frac{1}{3}\rho$. If we assume there is no energy loss ($\alpha=0$), then we obtain $p<-\frac{2}{3}\rho$ which is in agreement with the result obtained in \cite{MaartensWands2000} for the high energy regime.

With the usual expressions for the energy density and pressure of the inflaton scalar field,
\begin{equation}
\rho= \frac{1}{2}\dot\phi^2 +V 
\end{equation}
\begin{equation}
p= \frac{1}{2}\dot\phi^2 -V,
\end{equation}
the nonconservation of energy equation becomes:
\begin{equation}
\dot \phi(\ddot \phi+3H\dot \phi+V^\prime )=-\alpha \kappa^{18}_5(\frac{1}{2}\dot \phi^2 +V)^8.
\end{equation}
To obtain a negative pressure fluid, we employ the slow-roll mechanism. It is assumed that the $\ddot{\phi}$ and $\frac{1}{2}\dot{\phi}^2$ terms are negligible compared to the remaining terms. We thus obtain the following equation for $\phi$ 
\begin{equation}
3H\dot{\phi}^2+V^\prime \dot{\phi}+\alpha \kappa^{18}_5 V^8=0.
\end{equation}
This quadratic equation in $\dot{\phi}$ can be solved to give
\begin{equation}
\dot \phi= -\frac{V'}{6H}\pm  \frac{V^\prime}{6H}\sqrt {1-  \frac{12\alpha \kappa_5^{18} V^8H}{{V^\prime}^2}}.
\end{equation}
Due to the smallness of the loss term, the second term in the squareroot is much smaller than the first. The squareroot is therefore expanded to $O(\alpha)$ to give the two following solutions,
\begin{equation}
\dot{\phi}=\frac{-V^\prime}{3H}+\frac{\alpha\kappa_5^{18} V^8}{V^\prime}
\end{equation}
which essentially is just the (slightly modified) standard slow-roll dynamics, and
\begin{equation}
\dot{\phi}=-\frac{\alpha\kappa_5^{18} V^8}{V^\prime},
\end{equation}
which we dub the nonstandard inflationary branch.

So it is clear that the emission of gravitons (which causes the brane energy-momentum to not be conserved) has led to the possibility of having \emph{two different branches} of inflationary dynamics: the modified slow-roll branch and the nonstandard inflationary branch.

\subsection{Is the Nonstandard Inflationary Branch a Stable Attractor Solution?}

We would like to determine whether the nonstandard branch is stable to small perturbations. If it is, then we expect there to be a basin of attraction for this branch. The dynamics of the nonstandard inflationary branch should therefore be observed for a range of initial conditions. Or, in other words, the nonstandard branch will thus be a stable attractor solution.

To do this, we consider an inflationary solution starting slightly away from the nonstandard branch, $\phi_+$ and check analytically whether the perturbed solution converges to the nonstandard branch. For the purpose of this calculation, we specify to potentials of the form,
\begin{equation}
V=V(\phi)=\beta\phi^n.
\end{equation}
Writing $\phi$ as
\begin{equation}
 \phi=\phi_++\delta\phi,
\end{equation}
the equation for $\phi$
\begin{equation}
 \dot \phi(\ddot \phi+3H\dot \phi+V^\prime )=-\alpha \kappa^{18}_5 (\frac{1}{2}\dot \phi^2 +V)^8
\end{equation}
becomes
\begin{equation}
 (\dot \phi_++\dot {\delta\phi})(\ddot \phi_++\ddot {\delta\phi}+3H(\dot \phi_++\dot {\delta\phi}) + V^\prime)=-\alpha \kappa^{18}_5 (\frac{1}{2}(\dot \phi_++\dot {\delta\phi})^2 +V)^8
\end{equation}
We neglect second order terms, ${\delta\phi}^2, \delta\phi\dot {\delta\phi}$, etc. Also, we are only interested in the solution up till the end of inflation, so we assume that the slow-roll conditions hold for the nonstandard branch. We further define $V_+=V(\phi_+)=\beta\phi_+^n$ and $H_+=\frac{\kappa_5^2}{6}\left( \frac{1}{2}\dot \phi_+^2+V_+\right)$. The previous equation then reduces to:
\begin{equation}
 \dot \phi_+\ddot {\delta\phi}+3H\dot \phi_+^2+3H\dot \phi_+\dot {\delta\phi} +\dot \phi_+V^\prime+3H\dot {\delta\phi}\dot \phi_++\dot {\delta\phi}V^\prime=-\alpha \kappa^{18}_5 (8\dot \phi_+\dot {\delta\phi}V^7+V^8)
\end{equation}
We now need to determine $V$, $V^\prime$, and $H$,
\begin{equation}
 V=\beta\phi^n=\beta(\phi_++\delta\phi)^n=\beta\phi_+^n+{\beta}n\phi_+^{n-1}\delta\phi=V_++{\beta}n\phi_+^{n-1}\delta\phi
\end{equation}
\begin{equation}
 V^\prime=V_+^\prime+n(n-1)\beta\phi_+^{n-2}\delta\phi
\end{equation}
\begin{eqnarray}
 H=\frac{\kappa_5^2}{6}(\frac{1}{2}\dot \phi^2+V)=\frac{\kappa_5^2}{6}(\frac{1}{2}(\dot \phi_++\dot {\delta\phi})^2+V)=\frac{\kappa_5^2}{6}(\frac{1}{2}\dot \phi_+^2+\dot \phi_+\dot {\delta\phi}+V_++{\beta}n\phi_+^{n-1}\delta\phi)\nonumber \\
=H_++\frac{\kappa_5^2}{6}n\beta\phi_+^{n-1}\delta\phi+\frac{\kappa_5^2}{6}\dot \phi_+\dot {\delta\phi}.
\end{eqnarray}
Substituting these into the previous equation and neglecting second order terms gives:
\begin{equation}
\dot \phi_+\ddot {\delta\phi} + 3H_+\dot \phi_+^2 + \frac{\kappa_5^2}{2}n\beta\phi_+^{n-1}\dot \phi_+^2\delta\phi + \frac{\kappa_5^2}{2}\dot \phi_+^3\dot {\delta\phi} + 6H_+\dot \phi_+\dot {\delta\phi} + \dot \phi_+V_+^\prime + \dot \phi_+n(n-1)\beta\phi_+^{n-2}\delta\phi + \dot {\delta\phi}V_+^\prime $$
$$
= -\alpha\kappa_5^{18} \left( 8\dot \phi_+\dot {\delta\phi}V_+^7 + V_+^8 + 8V_+^7 V_+^{\prime}\delta\phi\right).
\end{equation}
For the nonstandard branch, we have that $3H_+\dot \phi_+^2+V_+^\prime\dot \phi_++\alpha\kappa_5^{18}V_+^8=0$ and that $\dot \phi_+=-\frac{\alpha\kappa_5^{18}V_+^8}{V_+^\prime}$.
Using the first of these equations, the equation for $\delta\phi$ reduces further to:
\begin{equation}
 \dot \phi_+\ddot {\delta\phi} + \frac{\kappa_5^2}{2}n\beta\phi_+^{n-1}\dot \phi_+^2\delta\phi + \kappa_5^2V_+\dot \phi_+\dot {\delta\phi} + \dot \phi_+n(n-1)\beta\phi_+^{n-2}\delta\phi + \dot {\delta\phi}V_+^\prime $$
$$
= -\alpha\kappa_5^{18} \left( 8\dot \phi_+\dot {\delta\phi}V_+^7 + 8V_+^7 V_+^{\prime}\delta\phi\right).
\end{equation}
Using $V_+=\beta\phi_+^n$ and $V_+^\prime=n\beta\phi_+^{n-1}$, and after some algebra, we obtain:
\begin{equation}
 \ddot {\delta\phi}+\left( \kappa_5^2\beta\phi_+^n + 8\alpha\kappa_5^{18}\beta^7\phi_+^{7n}+\frac{\beta n\phi_+^{n-1}}{\dot \phi_+}\right) \dot {\delta\phi}$$
$$
+\left( \frac{\kappa_5^2}{2}\beta n\phi_+^{n-1}\dot \phi_++n(n-1)\beta\phi_+^{n-2}+\frac{8\alpha\kappa_5^{18}n\beta^8\phi_+^{8n-1}}{\dot \phi_+}\right) \delta\phi=0.
\end{equation}
Substituting for $\dot \phi_+$ finally gives us:
\begin{equation}
 \ddot {\delta\phi}+\left( \kappa_5^2\beta\phi_+^n +\alpha\kappa_5^{18}\beta^7\phi_+^{7n} - \frac{n^2}{\alpha\kappa_5^{18}\beta^6\phi_+^{6n+2}}\right) \dot {\delta\phi}+\left( -\frac{\kappa_5^{20}}{2}\alpha\beta^8\phi_+^{8n}-n(7n+1)\beta\phi_+^{n-2}\right) \delta\phi=0.
\end{equation}
We now need to solve this 2nd order ODE to determine the behaviour of $\delta\phi$. Since we are only interested in whether $\phi$ converges to (or diverges from) $\phi_+$, we assume the slow-roll conditions for $\phi_+$ hold (otherwise $\phi_+$ would no longer be a valid solution). With this in mind, we treat $\phi_+$ as roughly constant, and so we just have an ODE with constant coefficients.

And so we have solutions of the form $e^{\lambda t}$ where $\lambda$ is either negative or very close to zero. We therefore see that we have a decaying solution and another that is almost constant (which just corresponds to a shift along the inflaton trajectory). Hence, the nonstandard inflationary branch is a stable attractor solution.

\subsection{Why is there a Second Branch?}

In this subsection we consider the origin of the second inflationary solution. It is straightforward to understand why the modified slow-roll branch exists; the energy loss term perturbs the equation of motion for $\phi$ slightly which then slightly perturbs this inflationary solution away from the standard slow-roll dynamics. But why is a second solution introduced?

In actual fact, the canonical equation of motion for the inflaton (i.e. with no energy loss) 
\begin{equation}
 \dot{\phi}\left(\ddot{\phi}+3H\dot{\phi} + V^{\prime}\right) = 0,
\end{equation}
has two solutions,
\begin{equation}
 \ddot{\phi}+3H\dot{\phi} + V^{\prime}=0,
\end{equation}
which reduces to the standard slow-roll dynamics if we assume the slow-roll conditions hold, and
\begin{equation}
 \dot{\phi}=0.
\end{equation}
This second solution is typically neglected because if $\phi$ is constant then there is no dynamical way of ending inflation and the universe would stay in a de Sitter phase indefinitely. However, the presence of energy loss perturbs this solution away from a constant $\phi$ solution, to give:
\begin{equation}
\dot{\phi}=-\frac{\alpha\kappa_5^{18} V^8}{V^\prime}.
\end{equation}
Thus, with energy loss, a second inflationary solution is allowed even for the canonical inflaton action.

\section{Calculation of the Spectrum of Scalar Perturbations}

\subsection{Curvature Perturbation Not Conserved}
The standard way of relating the fluctuations during inflation to late time observables is to calculate the power spectrum of quantum fluctuations at horizon exit and to conclude that this is equal to the power spectrum at horizon reentry at late times if we assume that the curvature perturbation is conserved on superhorizon scales (for the adiabatic perturbations which are generated in single-field slow roll inflation).

However, this is no longer the case when energy is not conserved on the brane (due to graviton emission, for example) and we therefore need to resort to different methods. The evolution of the perturbations must be tracked on superhorizon scales until horizon reentry at late times. In this subsection, we demonstrate that the curvature perturbation is not conserved on superhorizon scales even for the adiabatic perturbations generated in single-field slow-roll inflation. In the following subsections, we go through in detail the method we have used to calculate the late time power spectrum.

To demonstrate that the curvature perturbation is not conserved, we work in the Newtonian gauge for simplicity. Thus we use the following form of the perturbed Friedmann metric on the brane:
\begin{equation}
ds^2=-(1+2\Phi)dt^2+a^2[(1-2\Psi)\delta_{ij}]dx^idx^j.
\end{equation}
Only the scalar perturbations are of interest to us for this analysis, so we neglect vector and tensor modes.
For the calculation of the time derivative of the curvature perturbation, we require the nonconservation of energy equation.  Starting from the nonconservation of energy-momentum on the brane, we act with $u^\nu$, a timelike vector of a comoving observer on the brane, to obtain nonconservation of energy:
\begin{equation}
u_\nu\nabla_\mu T^{\mu\nu}=-\alpha\kappa_5^{18}\rho^8.
\end{equation}
And if we are interested in just the first order perturbation terms on superhorizon scales, we then have\footnote{We are now working in momentum space, but for notational convenience, the $k$-dependence has been suppressed.}
\begin{equation}
\dot{\delta\rho}+3H(\delta\rho+\delta{p})-3(\bar{\rho}+\bar{p})\dot{\Psi}=-8\alpha\kappa^{18}_5\delta\rho\bar{\rho}^7 - \Phi\alpha\kappa_5^{18}\bar{\rho}^8.
\end{equation}
The last term on the right arises from the perturbation to $u_0$ in the term $u_\nu\nabla_\mu T^{\mu\nu}$.

The zeroth order equation (which gives the background nonconservation equation) will also be useful:
\begin{equation}
\dot{\bar{\rho}}+3H(\bar{\rho}+\bar{p})=-\alpha\kappa_5^{18}\bar{\rho}^8.
\end{equation}
We want to calculate the time derivative of $\zeta$ which is the gauge-invariant variable representing the curvature perturbation on uniform density hypersurfaces, $\zeta=-\Psi-\frac{H}{\dot{\bar\rho}}\delta\rho$.
\begin{equation}
\dot{\zeta}=-\dot{\Psi}+\frac{H\dot{\delta\rho}}{3H(\bar{\rho}+\bar{p})+\alpha\kappa_5^{18}{\bar{\rho}}^8}-\frac{\alpha\kappa_5^{20}\bar{\rho}^8\delta\rho}{6(3H(\bar{\rho}+\bar{p})+\alpha\kappa_5^{18} {\bar{\rho}}^8)}$$
$$
+\frac{3H^2\left(1+\frac{\dot{\bar{p}}}{\dot{\bar{\rho}}}\right)\delta\rho}{3H(\bar{\rho}+ \bar{p}) +\alpha\kappa_5^{18} {\bar{\rho}}^8} + \frac{8\alpha H\kappa_5^{18}\bar{\rho}^7\delta\rho}{3H(\bar{\rho}+\bar{p})+\alpha\kappa_5^{18}\bar{\rho}^8}.
\end{equation}
The nonconservation of energy equation can also be rewritten as:
\begin{equation}
\frac{H\dot{\delta\rho}}{3H(\bar{\rho}+\bar{p})+\alpha\kappa_5^{18}\bar{\rho}^8}+\frac{3H^2\delta\rho}{3H(\bar{\rho}+\bar{p})+\alpha\kappa_5^{18}\bar{\rho}^8}+\frac{3H^2\delta p}{3H(\bar{\rho}+\bar{p})+\alpha\kappa_5^{18}\bar{\rho}^8}$$
$$ -\frac{3H(\bar{\rho}+\bar{p})\dot{\Psi}}{3H(\bar{\rho}+\bar{p})+\alpha\kappa_5^{18}\bar{\rho}^8}=-\frac{8\alpha\kappa^{18}_5H\delta\rho\bar{\rho}^7}{3H(\bar{\rho}+\bar{p})+\alpha\kappa_5^{18}\bar{\rho}^8} - \frac{\Phi H\alpha\kappa_5^{18}\bar{\rho}^8}{3H(\bar{\rho}+\bar{p})+\alpha\kappa_5^{18}\bar{\rho}^8}.
\end{equation}
We require the definition of the non-adiabatic pressure perturbation, $\delta{p_{nad}}$, as well,
\begin{equation}
\delta{p_{nad}}=\delta{p}-\frac{\dot{\bar{p}}\delta\rho}{\dot{\bar{\rho}}}.
\end{equation}
We can now combine the previous three equations to obtain:
\begin{equation}
\dot{\zeta}=-\frac{3H^2\delta{p_{nad}}}{3H(\bar{\rho}+\bar{p})+\alpha\kappa_5^{18}\bar{\rho}^8}-\frac{\alpha\kappa_5^{18}\bar{\rho}^8\dot{\Psi}}{3H(\bar{\rho}+\bar{p})+\alpha\kappa_5^{18}\bar{\rho}^8}-\frac{\alpha\Phi\kappa_5^{20}\bar{\rho}^9}{3H(\bar{\rho}+\bar{p})+\alpha\kappa_5^{18}\bar{\rho}^8}.
\end{equation}
And for adiabatic perturbations on superhorizon scales, the first term vanishes, leaving:
\begin{equation}
\dot{\zeta}=-\frac{\alpha\kappa_5^{18}\bar{\rho}^8\dot{\Psi}}{3H(\bar{\rho}+\bar{p})+\alpha\kappa_5^{18}\bar{\rho}^8}-\frac{\alpha\Phi\kappa_5^{20}\bar{\rho}^9}{3H(\bar{\rho}+\bar{p})+\alpha\kappa_5^{18}\bar{\rho}^8}.
\end{equation}
Thus, we see that the curvature perturbation is \emph{not conserved} on large scales, even for the purely adiabatic perturbations generated in single field inflation.

\subsection{Quantisation of Inflaton Fluctuations}
In the previous subsection, we have demonstrated that the curvature perturbation is not conserved on superhorizon scales. The power spectrum at horizon reentry at late times therefore cannot merely just be matched to the power spectrum of quantum fluctuations at Hubble exit. New methods are required to track the superhorizon evolution of the curvature perturbation through till horizon reentry during the late time low energy regime. To achieve this, we adopt a method commonly used in multifield slow-roll inflation models (where here the non-adiabatic perturbations cause the curvature perturbation to evolve on superhorizon scales). Essentially, the method is as follows: the perturbation equations are solved on superhorizon scales, then, to fix the integration constants that arise, the amplitude of the quantum fluctuations at horizon exit is used as an initial condition. We now have a solution to the perturbation equations that is valid on superHubble scales, so we can use this to evaluate the power spectrum at horizon reentry at late times.

In actual fact, our case is slightly more complicated than this due to the presence of the two different regimes: the high energy regime and the low energy regime. More specifically, the perturbation equations must first be solved on superhorizon scales in the high energy regime under slow-roll inflationary conditions and assuming that the adiabatic mode is weakly time-varying. The integration constants which arise are then fixed by comparison to the power spectrum of quantum fluctuations at Hubble exit. We assume that this period of high energy slow-roll inflation is followed by a low energy radiation era in which standard 4D general relativity holds and standard cosmology is recovered. And once we have recovered standard 4D general relativity, we know that the curvature perturbation $\zeta$ will be conserved on superhorizon scales (for the adiabatic perturbations in single-field inflation)\cite{Weinberg2003, Wands2000}. So the superhorizon solution found previously which is valid only during the high energy slow-roll inflationary period can be used to calculate the value of $\zeta$ at the end of inflation, which will then be conserved until horizon reentry at late times. The power spectrum of $\zeta$ at late time horizon reentry can thus easily be obtained.

In principle, when performing this calculation, the full system of bulk and brane metric perturbations together with the inflaton perturbation should be considered in its entirety. We cannot just perturb about a 4D Friedmann metric but should include the coupling to the bulk metric perturbation as well. This complicates our calculation in two ways. First, to obtain the power spectrum of scalar perturbations at horizon exit, we should quantise the complete set of perturbations for coupled brane-bulk system and not merely perturb about a 4D Friedmann metric and neglect bulk effects. However, it was shown in \cite{Rubakov2007} that even when the coupled system of brane-bulk scalar perturbations is consistently quantised, the bulk metric perturbation only introduces corrections at first order in the slow-roll parameters. So, in this instance, we will neglect the bulk metric perturbation, and quantise the system of inflaton-brane metric perturbations only. Admittedly we are using a different bulk metric to \cite{Rubakov2007}, but we assume that any additional corrections introduced by our bulk metric will be further suppressed by $\alpha$, the coefficient of our loss term, and this simplifying assumption should therefore remain valid.

The second way in which the coupling to the bulk metric perturbation affects our calculation is its effect on the superhorizon evolution of the brane scalar perturbations during the high energy inflationary regime. When considering how the scalar perturbations on the brane evolve, we should again consider the full brane-bulk system of equations. However, it was shown in \cite{Wands2000} that for any braneworld model where the 4D energy-momentum tensor is conserved, then the brane curvature perturbation is conserved on superhorizon scales (for pure adiabatic brane perturbations). So for our model, we assume that it is the nonconservation of energy on the brane and not the higher-dimensional bulk perturbations which is the dominant effect on the superhorizon evolution of the brane scalar perturbations. We therefore neglect the effect of the bulk perturbations for the rest of this section.

In this subsection, we aim to quantise the brane scalar fluctuations to determine the power spectrum at Hubble exit. There are two ways in which we could proceed. First, we could start from the second order action, change variable appropriately, and quantise. However, in our case, while we could start from a 5D action, we have made several simplifying assumptions regarding the bulk effects, so perhaps this is not the best way to proceed. Alternatively, we could start from the scalar perturbation equations, define a Mukhanov variable \cite{Mukhanov1985} such that we obtain the harmonic oscillator equation, and then quantise. By right, since the equations we obtain are linear, any new variable defined through some numerical rescaling would be a viable solution as well. Thus, to ensure that we have chosen the right normalisation, we demand that, in the $\alpha\rightarrow 0$ limit (where standard inflationary theory is recovered), the standard solution to the mode equation is reobtained.

We first derive the perturbation equation for $\delta\phi$\footnote{Our method in this subsection is similar to \cite{PolarskiStarobinsky}.}. We start with the nonconservation of energy equation,
\begin{equation}
u_\nu\nabla_\mu T^{\mu\nu}=-\alpha\kappa_5^{18}\rho^8.
\end{equation}
and then focus on the first order perturbation terms which gives the equation
\begin{equation}
\dot{\delta\rho}+3H(\delta\rho+\delta p)-3\dot{\Phi}(\bar{\rho}+\bar{p}) - \frac{k^2}{a^2}\delta q=-8\alpha\kappa_5^{18}\delta\rho\bar{\rho}^7-\Phi\alpha\kappa_5^{18}\bar{\rho}^8,
\end{equation}
in the Newtonian gauge and for zero anisotropic stress (which is true if we neglect bulk effects). 

Using the expressions for $\delta\rho$, $\delta p$, and $\delta q$ in slow-roll inflation
\begin{equation}
\delta\rho=\dot{\phi}\dot{\delta\phi}-\Phi\dot{\phi}^2+V^{\prime}\delta\phi
\end{equation}
\begin{equation}
\delta p=\dot{\phi}\dot{\delta\phi}-\Phi\dot{\phi}^2-V^{\prime}\delta\phi,
\end{equation}
\begin{equation}
 \delta q=-\dot{\phi}\delta\phi
\end{equation}
the perturbation equation becomes
\begin{equation}
\dot{\phi}\ddot{\delta\phi}+\ddot{\phi}\dot{\delta\phi}-4\dot{\Phi}\dot{\phi}^2-2\Phi\dot{\phi}\ddot{\phi}+V^{\prime\prime}\dot{\phi}\delta\phi +V^{\prime}\dot{\delta\phi}+6H\dot{\phi}\dot{\delta\phi}-6H\Phi\dot{\phi}^2 +\frac{k^2}{a^2}\dot{\phi}\delta\phi$$
$$
=-8\alpha\kappa_5^{18}(\dot{\phi}\dot{\delta\phi}-\Phi\dot{\phi}^2+V^{\prime}\delta\phi)\left(\frac{1}{2}\dot{\phi}^2+V\right)^7-\Phi\alpha\kappa_5^{18}\left(\frac{1}{2}\dot{\phi}^2+V\right)^8.
\end{equation}
This equation can be simplified slightly by using the background equation,
\begin{equation}
\dot{\phi}(\ddot{\phi}+3H\dot{\phi}+V^{\prime})=-\alpha\kappa_5^{18}\left(\frac{1}{2}\dot{\phi}^2+V\right)^8
\end{equation}
 to remove the second time derivatives of $\phi$. It simplifies to
\begin{equation}
\ddot{\delta\phi}+3H\dot{\delta\phi}-4\dot{\Phi}\dot{\phi}+V^{\prime\prime}\delta\phi+2V^{\prime}\Phi+\frac{k^2}{a^2}\delta\phi$$
$$
=\alpha\kappa_5^{18}\left(\frac{1}{2}\dot{\phi}^2+V\right)^7\left[\frac{13}{2}\Phi\dot{\phi}-\frac{3\Phi V}{\dot{\phi}}-\frac{15}{2}\dot{\delta\phi}+\frac{\dot{\delta\phi} V}{\dot{\phi}^2}-\frac{8V^{\prime}\delta\phi}{\dot{\phi}}\right],
\end{equation}
which is our starting point.

In the extreme slow-roll limit, all mass- and $\Phi$-dependent terms (which are $O(\epsilon, \eta)$) may be neglected\footnote{This is equivalent to using the pure de Sitter result at horizon exit when calculating the power spectrum of quantum fluctuations.}. The previous equation reduces to:
\begin{equation}
\ddot{\delta\phi}+3H\dot{\delta\phi}+\frac{k^2}{a^2}\delta\phi=\alpha\kappa_5^{18}V^7\left[\frac{V\dot{\delta\phi}}{\dot{\phi}^2}-\frac{8V^{\prime}\delta\phi}{\dot{\phi}}\right].
\end{equation}
Changing the independent variable to conformal time gives:
\begin{equation}
{\delta\phi}^{\prime\prime}+\left[\frac{2{a}^\prime}{a}-\frac{\alpha\kappa_5^{18} V^8 a}{\dot{\phi}^2}\right]{\delta\phi}^\prime +\left[k^2 + \frac{8\alpha\kappa_5^{18} V^7 V^{\prime} a^2}{\dot{\phi}}\right] \delta\phi=0.
\end{equation}
We now wish to solve this equation. The standard strategy is to change variable (to the Mukhanov variable) for which the equation changes to the canonical form of the harmonic oscillator equation, which can then easily be quantised. If we are given an equation of the form

\begin{equation}
{\delta\phi}^{\prime\prime}+A{\delta\phi}^{\prime}+B\delta\phi=0,
\end{equation}
we introduce the Mukhanov variable, $v=z\delta\phi$, such that the equation for $v$ will have no first derivative (so essentially, we will then have the harmonic oscillator equation). The condition for this is simply:

\begin{equation}
\frac{2z^\prime}{z}=A.
\end{equation}
With this in mind, we define
\begin{equation}
z=ae^{-\int\frac{\alpha\kappa_5^{18} V^8a}{2\dot{\phi}^2}} ; \text{where } v=z\delta\phi.
\end{equation}
We now have the following equation
\begin{equation}
v^{\prime\prime}+\left[k^2+\frac{8\alpha\kappa_5^{18} V^7 V^{\prime}a^2}{\dot{\phi}}-\frac{z^{\prime\prime}}{z}\right]v=0.
\end{equation}
The solution to this should be proportional to the solution of the mode equation for canonical inflation, up to $O(\alpha)$ corrections, that is:
\begin{equation}
v\propto\frac{e^{ik\eta}}{\sqrt{2k}}\left(1-\frac{i}{k\eta}\right) + O(\alpha).
\end{equation}
The normalisation of this solution is therefore still arbitrary. We fix the normalisation by demanding that the canonically normalised solution is obtained in the $\alpha \rightarrow 0$ limit. Thus,
\begin{equation}
v=\frac{e^{ik\eta}}{\sqrt{2k}}\left(1-\frac{i}{k\eta}\right) + O(\alpha).
\end{equation}
Now it is not easy to write down an exact analytic expression for the  $O(\alpha)$ term, so to make further progress analytically, we use the approximate $\alpha=0$ solution for $v$ (acknowledging that this will introduce an error into our final result).

Following the standard procedure, we obtain the following power spectrum (evaluated at horizon exit)
\begin{eqnarray}
\langle{\delta\phi}_{k}{\delta\phi}_{k^{\prime}}\rangle &=(2\pi)^3\delta(k+k^\prime)\frac{|v^2|}{z^2} \nonumber
\\ &=(2\pi)^2\delta(k+k^{\prime})\frac{H^2}{2k^3}e^{\int\frac{\alpha\kappa_5^{18} V^8 k}{\dot{\phi}^2 H}}
\end{eqnarray}
So, the dimensionless and dimensionful power spectra ($\Delta^2_{\delta\phi} \text{and }P_{\delta\phi}$ respectively) are given by
\begin{equation}
\Delta^2_{\delta\phi}=\frac{k^3}{2\pi^2}P_{\delta\phi}=\frac{H^2}{(2\pi)^2}e^{\int\frac{\alpha\kappa_5^{18} V^8 k}{\dot{\phi}^2 H}}.
\end{equation}
Surprisingly, there is an \textit{explicit} dependence on scale, $k$.

\subsection{Solution of $\Phi$ During the High Energy Inflationary Period}

We now wish to solve the perturbation equations during the high energy inflationary regime on superhorizon scales. The perturbation equation for $\delta\phi$ on superhorizon scales is given by,
\begin{equation}
\ddot{\delta\phi}+3H\dot{\delta\phi}-4\dot{\Phi}\dot{\phi}+V^{\prime\prime}\delta\phi+2V^{\prime}\Phi+\frac{k^2}{a^2}\delta\phi$$
$$
=\alpha\kappa_5^{18}\left(\frac{1}{2}\dot{\phi}^2+V\right)^7\left[\frac{13}{2}\Phi\dot{\phi}-\frac{3\Phi V}{\dot{\phi}}-\frac{15}{2}\dot{\delta\phi}+\frac{\dot{\delta\phi} V}{\dot{\phi}^2}-\frac{8V^{\prime}\delta\phi}{\dot{\phi}}\right],
\end{equation}
We want to solve this equation under slow-roll conditions and assuming that the adiabatic mode is weakly time-dependent so that we can neglect the time derivative of $\Phi$ as well as second time derivatives. The equation simplifies further to
\begin{equation}
3H\dot{\delta\phi}+V^{\prime\prime}\delta\phi+2V^{\prime}\Phi=\alpha\kappa_5^{18} V^7\left[\frac{\dot{\delta\phi}V}{\dot{\phi}^2}-\frac{3\Phi V}{\dot{\phi}}-\frac{8V^{\prime}\delta\phi}{\dot{\phi}}\right].
\end{equation}
Note that unlike in the previous subsection, not all the $\Phi$-dependent terms disappear, hence we combine this equation with the $0i$-component of the Einstein equations to obtain a closed set of equations.

The $0i$-component of the Einstein equations is given by \footnote{By right, there should be additional terms related to the bulk on the RHS but as mentioned earlier, we choose to neglect bulk effects since we assume that nonconservation of energy will be the dominant effect.}
\begin{equation}
H\Phi+\dot{\Phi}=\dot{\phi}\delta\phi,
\end{equation}
which under the assumption that $\dot{\Phi}$ is negligible reduces to
\begin{equation}
H\Phi=\dot{\phi}\delta\phi.
\end{equation}
We solve this coupled set of equations using the method in \cite{Starobinsky1995}. We first substitute the Einstein $0i$-equation into the first order perturbation equation of energy nonconservation, which results in:
\begin{equation}
\left[3H-\frac{\alpha\kappa_5^{18}V^8}{\dot{\phi}^2}\right]\dot{\delta\phi}+\left[V^{\prime\prime}+\frac{2V^{\prime}\dot{\phi}}{H}+\frac{3\alpha\kappa_5^{18}V^8}{H}+\frac{8\alpha\kappa_5^{18} V^7 V^{\prime}}{\dot{\phi}}\right]\delta\phi=0.
\end{equation}
 The independent variable is then changed from time to $\phi$ (that is, $\dot{\delta\phi}={\delta\phi}^\prime \dot{\phi}$ where $\prime$ denotes differentiation wrt to $\phi$).
\begin{equation}
\left[3H\dot{\phi}-\frac{\alpha\kappa_5^{18}V^8}{\dot{\phi}}\right]\dot{\delta\phi}+\left[V^{\prime\prime}+\frac{2V^{\prime}\dot{\phi}}{H}+\frac{3\alpha\kappa_5^{18}V^8}{H}+\frac{8\alpha\kappa_5^{18} V^7 V^{\prime}}{\dot{\phi}}\right]\delta\phi=0.
\end{equation}
Before continuing to solve these equations, we first check to see if this result agrees with standard low energy inflation (i.e. the case with $\alpha=0$ and $H^2=\frac{2}{3}V$). Under these conditions, our previous equation reduces to just
\begin{equation}
{\delta\phi}^{\prime}+\left[\frac{V^{\prime\prime}}{3H\dot{\phi}}+\frac{2V^{\prime}}{3H^2}\right]\delta\phi=0.
\end{equation}
And using standard inflationary dynamics $3H\dot{\phi}=-V^\prime$, this becomes
\begin{equation}
{\delta\phi}^{\prime}+\left[\frac{V^{\prime}}{V}-\frac{V^{\prime\prime}}{V^{\prime}}\right]\delta\phi=0.
\end{equation}
This first order differential equation can easily be integrated to obtain
\begin{equation}
\delta\phi=\frac{AV^\prime}{V}, \text{$A$ some constant}
\end{equation}
which is the correct result for standard low energy inflation \cite{MukhanovSteinhardt1997}.

We now solve the perturbation equations for our inflationary dynamics, i.e. both for the modified slow-roll branch and the nonstandard inflationary branch.

The background dynamics for the modified slow-roll branch are given by
\begin{equation}
\dot{\phi}=\frac{-V^\prime}{3H}+\frac{\alpha\kappa_5^{18} V^8}{V^\prime}
\end{equation}
With this and the high energy result for the Friedmann equation ($H=\frac{\kappa_5^2 V}{6}$), the perturbation equation becomes (to leading order in $\alpha$)
\begin{equation}
\left[-V^\prime+\frac{6\alpha\kappa_5^{18} V^8H}{V^\prime}\right]\delta\phi^\prime+\left[V^{\prime\prime}-\frac{2{V^{\prime}}^2}{3H^2}+\frac{5\alpha\kappa_5^{18}V^8}{H}  - 24\alpha \kappa_5^{18} V^7H\right]\delta\phi=0.
\end{equation}
Dividing through by the coefficient of the first term, we obtain
\begin{equation}
\delta\phi^\prime+\left[\frac{-V^{\prime\prime}}{V^\prime}+\frac{2V^{\prime}}{3H^2}+\alpha L\right]\delta\phi=0,
\end{equation}
where $\alpha L=-\frac{\alpha\kappa_5^{18}V^8}{HV^\prime}+\frac{24\alpha\kappa_5^{18} V^7H}{V^\prime}-\frac{6\alpha\kappa_5^{18} V^8H V^{\prime\prime}}{{V^\prime}^3}$.

Again, this first order differential equation can be integrated to obtain:
\begin{equation}
\delta\phi=C V^\prime e^{\frac{24}{\kappa_5^4 V}}e^{-\int \alpha L}, \text{some constant $C$}.
\end{equation}
And for the nonstandard branch, the background dynamics are given by
\begin{equation}
\dot{\phi}=-\frac{\alpha\kappa_5^{18} V^8}{V^\prime}.
\end{equation}
To leading order in $\alpha$, the perturbation equation becomes
\begin{equation}
\delta\phi^\prime+\left[\frac{V^{\prime\prime}}{V^\prime}-\frac{8V^{\prime}}{V}+\alpha M\right]\delta\phi=0.
\end{equation}
This can be solved to obtain
\begin{equation}
\delta\phi=D\frac{V^8}{V^\prime}e^{-\int\alpha M}, \text{ some constant $D$}
\end{equation}
where $\alpha M=\frac{\alpha\kappa_5^{18} V^8}{HV^\prime}+\frac{3\alpha\kappa_5^{18} V^8 V^{\prime\prime}}{{V^\prime}^2}+\frac{24H\alpha\kappa_5^{18} V^7}{V^\prime}$.

The constants C and D from the solutions of $\delta\phi$ on superhorizon scales in the high energy inflationary regime can now be determined by comparison to the power spectrum of $\delta\phi$ at horizon exit. We have the following expressions for $\delta\phi$ for the modified slow-roll branch and the nonstandard branch respectively

\begin{eqnarray*}
\delta\phi&=CV^\prime e^{\frac{24}{\kappa_5^4 V}}e^{-\int \alpha L}\\
\delta\phi&=D\frac{V^8}{V^\prime}e^{-\int \alpha M},
\end{eqnarray*}
Rearranging gives
\begin{eqnarray}
C&=\Big[\frac{\delta\phi}{V^\prime} e^{-\frac{24}{\kappa_5^4 V}}e^{\int \alpha L}\Big]_{k=aH}\\
D&=\Big[\frac{\delta\phi V^\prime}{V^8}e^{\int \alpha M}\Big]_{k=aH},
\end{eqnarray}
where, at horizon crossing, the quantum fluctuations of $\delta\phi$ are given by
\begin{equation}
\delta\phi\sim\frac{H}{(2\pi)}e^{\int\frac{\alpha\kappa_5^{18} V^8 k}{2\dot{\phi}^2 H}}.\nonumber
\end{equation}
So the high energy inflationary solutions for $\delta\phi$ on superhorizon scales for both the modified slow roll branch and the nonstandard branch are given respectively by:

\begin{eqnarray}
\delta\phi&=\Big[\frac{\delta\phi}{V^\prime} e^{-\frac{24}{\kappa_5^4 V}}e^{\int \alpha L}\Big]_{k=aH}V^\prime e^{\frac{24}{\kappa_5^4 V}}e^{-\int \alpha L}\\
\delta\phi&=\Big[\frac{\delta\phi V^\prime}{V^8}e^{\int \alpha M}\Big]_{k=aH}\frac{V^8}{V^\prime}e^{-\int \alpha M}.
\end{eqnarray}
And using the relation for $\Phi$
\begin{equation}
H\Phi=\dot{\phi}\delta\phi,\nonumber
\end{equation}
we find the following expressions for $\Phi$ in the modified slow roll and nonstandard branches respectively,
\begin{eqnarray}
\Phi&=\Big[\frac{\delta\phi}{V^\prime} e^{-\frac{24}{\kappa_5^4 V}}e^{\int \alpha L}\Big]_{k=aH}\frac{\dot{\phi}V^\prime e^{\frac{24}{\kappa_5^4 V}}e^{-\int \alpha L}}{H}\\
\Phi&=\Big[\frac{\delta\phi V^\prime}{V^8}e^{\int \alpha M}\Big]_{k=aH}\frac{\dot{\phi}V^8}{H V^\prime}e^{-\int \alpha M}.
\end{eqnarray}

\subsection{Matching the Solution to the Low Energy Era}
It must be stressed that these solutions are only valid on superhorizon scales during the period of high energy slow-roll inflation. However, we are interested in tracking these solutions until the modes reenter the Hubble radius at late times. As mentioned earlier, we assume that the period of high energy inflation is followed by a low energy radiation era in which standard 4D general relativity holds. And once standard general relativity is recovered we know that $\zeta$, the curvature perturbation on uniform density hypersurfaces, will be conserved on superhorizon scales \cite{Wands2000, Weinberg2003}. We therefore evaluate $\zeta$ at the end of inflation (which takes place at the High Energy-Low Energy Transition (HE-LE)) using the superhorizon solutions found in the previous subsection. $\zeta$ will then remain constant on superhorizon scales, therefore the expression for $\zeta$ at the HE-LE transition will be valid until horizon reentry at late times.

We will now evaluate $\zeta$ at the HE-LE transition using the superhorizon solutions that we found previously.

$\zeta$ is given by the following expression
\begin{equation}
-\zeta=\Phi + \frac{H\delta\rho}{\dot{\bar{\rho}}},
\end{equation}
which during inflation can be expressed as
\begin{equation}
 -\zeta=\Phi+\frac{H\delta\phi}{\dot{\phi}}+O(\alpha^2).
\end{equation}
Using the solutions for $\Phi$ and $\delta\phi$, we obtain the following results:
\begin{eqnarray}
\zeta &= \Big[\frac{1}{V^\prime} e^{-\frac{24}{\kappa_5^4 V}}e^{\int \alpha L}\Big]_{k=aH} \Big[V^\prime e^{\frac{24}{\kappa_5^4 V}}e^{-\int \alpha L} \Big]_{HE-LE} \left[\frac{\dot{\phi}}{H}+\frac{H}{\dot{\phi}}\right]_{HE-LE}\left[\delta\phi\right]_{k=aH}\nonumber \\
&\approx \Big[\frac{1}{V^\prime} e^{-\frac{24}{\kappa_5^4 V}}e^{\int \alpha L}\Big]_{k=aH} \Big[V^\prime e^{\frac{24}{\kappa_5^4 V}}e^{-\int \alpha L} \Big]_{HE-LE} \left[\frac{H}{\dot{\phi}}\right]_{HE-LE}\left[\delta\phi\right]_{k=aH}.
\end{eqnarray}
if inflation proceeded along the modified slow-roll branch, or
\begin{eqnarray}
\zeta&=\Big[\frac{V^\prime}{V^8}e^{\int \alpha M}\Big]_{k=aH}\Big[\frac{V^8}{V^\prime}e^{-\int \alpha M}\Big]_{HE-LE}\left[\frac{\dot{\phi}}{H}+\frac{H}{\dot{\phi}}\right]_{HE-LE}\left[\delta\phi\right]_{k=aH}\nonumber\\
&\approx\Big[\frac{V^\prime}{V^8}e^{\int \alpha M}\Big]_{k=aH}\Big[\frac{V^8}{V^\prime}e^{-\int \alpha M}\Big]_{HE-LE}\left[\frac{H}{\dot{\phi}}\right]_{HE-LE}\left[\delta\phi\right]_{k=aH}
\end{eqnarray}
if inflation followed the nonstandard inflationary dynamics.

\subsection{Calculation of Power Spectrum}
To obtain the power spectrum of $\zeta$ at horizon reentry, we simply have to calculate the power spectrum at the HE-LE transition since $\zeta$ will be conserved until horizon reentry. Using the results of the previous subsection, we therefore have:
\begin{eqnarray}
\Delta^2_{\zeta}&= \Big[\frac{1}{V^\prime} e^{-\frac{24}{\kappa_5^2 V}}e^{\int \alpha L}\Big]_{k=aH}^2 \Big[V^\prime e^{\frac{24}{\kappa_5^4 V}}e^{-\int \alpha L} \Big]_{HE-LE}^2 \left[\frac{\dot{\phi}}{H}+\frac{H}{\dot{\phi}}\right]^2_{HE-LE}\left[\Delta^2_{\delta\phi}\right]_{k=aH}\nonumber \\
&\approx \Big[\frac{1}{V^\prime} e^{-\frac{24}{\kappa_5^2 V}}e^{\int \alpha L}\Big]_{k=aH}^2 \Big[V^\prime e^{\frac{24}{\kappa_5^4 V}}e^{-\int \alpha L} \Big]_{HE-LE}^2 \left[\frac{H}{\dot{\phi}}\right]^2_{HE-LE}\left[\Delta^2_{\delta\phi}\right]_{k=aH}.
\end{eqnarray}
if inflation proceeded along the modified slow-roll branch, or
\begin{eqnarray}
\Delta^2_{\zeta}&=\Big[\frac{V^\prime}{V^8}e^{\int \alpha M}\Big]_{k=aH}^2\Big[\frac{V^8}{V^\prime}e^{-\int \alpha M}\Big]_{HE-LE}^2\left[\frac{\dot{\phi}}{H}+\frac{H}{\dot{\phi}}\right]^2_{HE-LE}\left[\Delta^2_{\delta\phi}\right]_{k=aH}\nonumber \\
&\approx\Big[\frac{V^\prime}{V^8}e^{\int \alpha M}\Big]_{k=aH}^2\Big[\frac{V^8}{V^\prime}e^{-\int \alpha M}\Big]_{HE-LE}^2\left[\frac{H}{\dot{\phi}}\right]^2_{HE-LE}\left[\Delta^2_{\delta\phi}\right]_{k=aH}
\end{eqnarray}
if inflation followed the nonstandard inflationary dynamics. Here, $\Delta^2_{\delta\phi}$ is the power spectrum of inflaton fluctuations at Hubble exit, which we obtained earlier as
\begin{equation}
\Delta^2_{\delta\phi}=\frac{k^3}{2\pi^2}P_{\delta\phi}=\frac{H^2}{(2\pi)^2}e^{\int\frac{\alpha\kappa_5^{18} V^8 k}{\dot{\phi}^2 H}}.
\end{equation}

\subsection{Calculation of Spectral Index}
In this section, we aim to calculate the spectral index given by the formula,
\begin{equation}
 n_s-1=\frac{d\log\Delta^2_{\zeta}}{d\log k}.
\end{equation}
We first need to calculate the slow-roll parameter, $\epsilon$, given by
\begin{equation}
 \epsilon\equiv - \frac{\dot{H}}{H^2} = -\frac{d\log H}{dN}.
\end{equation}
Using the nonconservation of energy equation
\begin{equation}
 \dot{\phi}(\ddot{\phi}+3H\dot{\phi}+V^{\prime})=-\alpha \kappa_5^{18} V^8,
\end{equation}
and the time derivative of the Friedmann equation
\begin{equation}
 \dot{H}=\frac{\kappa_5^2}{6}\left(\ddot{\phi}\dot{\phi} + V^{\prime}\dot{\phi}\right),
\end{equation}
we obtain the following expression,
\begin{equation}
 \epsilon=\frac{\kappa_5^2 \dot{\phi}^2}{2H} + \frac{\alpha\kappa_5^{20}V^8}{6H^2}.
\end{equation}
We can now use this to obtain expressions for $V^\prime$ for the modified slow-roll branch and the nonstandard branch, given by:
\begin{equation}
 {V^{\prime}}^2=\frac{18H^3\epsilon}{\kappa_5^2} + \alpha 6^8 3 \kappa_5^2 H^9
\end{equation}
and
\begin{equation}
 {V^{\prime}}^2=\frac{\kappa_5^6\alpha^26^{15} 3 H^{15}}{\epsilon}
\end{equation}
respectively.

And substituting these into the previous expressions for the power spectra give
\begin{equation}
 \Delta^2_{\zeta}\approx \left[\frac{1}{\left(\frac{18H^3\epsilon}{\kappa_5^2} + \alpha 6^8 3 \kappa_5^2 H^9\right)}e^{-\frac{8}{\kappa_5^2 H}} e^{2\int \alpha L}\right]_{k=aH} \left[\frac{H^2}{\left(2\pi\right)^2} e^{\int\frac{\alpha\kappa_5^{20} V^8k}{2H^2\left(\epsilon-\frac{\alpha\kappa_5^{20}V^8}{6H^2}\right)}}\right]_{k=aH}\left[\text{\textperiodcentered}\right]_{HE-LE}.
\end{equation}
for the modified slow-roll branch and
\begin{equation}
 \Delta^2_{\zeta}\approx \left[\frac{\kappa_5^6 \alpha^2 3^{16} 2^{15} H^{15}}{\epsilon} \frac{\kappa_5^{32}}{6^{16} H^{16}} e^{2\int \alpha M}\right]_{k=aH} \left[\frac{H^2}{\left(2\pi\right)^2}e^{\int\frac{\alpha\kappa_5^{20} V^8 k}{2H^2\left(\epsilon-\frac{\alpha\kappa_5^{20}V^8}{6H^2}\right)}}\right]_{k=aH}\left[\text{\textperiodcentered}\right]_{HE-LE}.
\end{equation}
for the nonstandard branch. The terms evaluated at the HE-LE transition have not been specified because they will not contribute to the spectral index as they do not have any scale-dependence.

If we act with\footnote{We use this expression since there is explicit $k$-dependence as well $k$-dependence implicit from the fact that different modes exit the horizon at different times.}
\begin{equation}
 \frac{d}{d \log k}=\frac{\partial}{\partial\log k}+\frac{dN}{d\log k}\frac{\partial}{\partial N}
\end{equation}
on these expressions for the power spectra, we then obtain
\begin{equation}
 n_s-1= -2\epsilon + \frac{-3\epsilon +\epsilon\eta - \epsilon \alpha 3^9 2^7\kappa_5^4 H^6}{\epsilon + \alpha 3^7 2^7 \kappa_5^4 H^6} - \frac{8\epsilon}{\kappa_5^2 H} + \frac{2\sqrt{2}\alpha L}{H^{\frac{1}{2}}\kappa_5^2}\left(\epsilon - \alpha 6^7 \kappa_5^4 H^6\right)^{\frac{1}{2}}$$
$$ + \frac{3^8 2^7 \alpha \kappa_5^4 H^6}{(\epsilon - 6^7\alpha\kappa_5^4H^6)} + \int \frac{3^8 2^7 \alpha \kappa_5^4 k H^6}{(\epsilon - 6^7\alpha\kappa_5^4H^6)},
\end{equation}
for the modified slow-roll branch, and
\begin{equation}
 n_s -1 = -\epsilon - \eta + \frac{2\sqrt{2}\alpha M}{H^{\frac{1}{2}}\kappa_5}\left(\epsilon - \alpha 6^7 \kappa_5^4 H^6\right)^{\frac{1}{2}} $$
$$
+ \frac{3^8 2^7 \alpha \kappa_5^4 H^6}{\left(\epsilon - 6^7 \alpha \kappa_5^4 H^6\right)} + \int \frac{3^8 2^7 \alpha \kappa_5^4 k H^6}{\left(\epsilon - 6^7 \alpha \kappa_5^4 H^6\right)}
\end{equation}
for the nonstandard branch, where $\eta=\frac{d\log \epsilon}{dN}$.

\subsection{The $\alpha \rightarrow 0$ limit}
We find that the power spectra and spectral indices that we obtain when there is energy loss from the brane are quite different from the standard expressions. For standard inflation in the high energy regime, we should obtain
\begin{equation}
\Delta^2_{\zeta}=\left[\frac{H^2}{\dot{\phi}^2}\frac{H^2}{(2\pi)^2}\right]_{k=aH}.
\end{equation}
for the power spectrum, and
\begin{equation}
 n_s-1= -3\epsilon - \eta.
\end{equation}
for the spectral index.

However, the power spectrum and spectral index obtained for the modified slow-roll branch should reduce to the standard expressions in the limit $\alpha\rightarrow 0$. It is easy to see why this is true for the power spectrum. In the limit $\alpha\rightarrow 0$, $\Delta^2_{\zeta}$ should reduce to its value at horizon exit, since if there is no energy loss from the brane, $\zeta$ must be conserved on superhorizon scales. All terms evaluated at the HE-LE transition should therefore be able to be expressed as their corresponding values at horizon exit ($k=aH$) plus $O(\alpha)$ corrections. The power spectrum for the modified slow-roll branch can thus be expressed as
\begin{equation}
 \Delta^2_{\zeta}= \Big[\frac{1}{V^\prime} e^{-\frac{24}{\kappa_5^2 V}}e^{\int \alpha L}\Big]_{k=aH}^2 \Big[V^\prime e^{\frac{24}{\kappa_5^4 V}}e^{-\int \alpha L} \Big]_{k=aH}^2 \left[\frac{H}{\dot{\phi}}\right]^2_{k=aH}\left[\Delta^2_{\delta\phi}\right]_{k=aH} + O(\alpha),
\end{equation}
which is essentially just
\begin{equation}
 \Delta^2_{\zeta}= \left[\frac{H}{\dot{\phi}}\right]^2_{k=aH}\left[\frac{H^2}{(2\pi)^2}e^{\int\frac{\alpha\kappa_5^{18} V^8 k}{\dot{\phi}^2 H}}\right]_{k=aH} + O(\alpha).
\end{equation}
And this clearly reduces to
\begin{equation}
 \Delta^2_{\zeta}=\left[\frac{H^2}{\dot{\phi}^2}\frac{H^2}{(2\pi)^2}\right]_{k=aH}
\end{equation}
if we take $\alpha\rightarrow 0$.

Note that this argument does not apply to the nonstandard branch because $\dot{\phi}\propto \alpha$ and the expression for the power spectrum will diverge in the limit $\alpha\rightarrow 0$. 

For the spectral index however, if one works directly from the expression for the spectral index of the modified slow-roll branch, it is less straightforward to recover the standard expression for high energy inflation. This is because energy loss, which results in superhorizon evolution beyond horizon exit, in a sense causes a loss of scale dependence because each mode must be evolved to and then evaluated at the same point in time (the HE-LE transition). In contrast, when there is no energy loss, each mode is evaluated when it crosses the horizon, and this introduces some terms in the expression for the spectral index which are not present for the case with energy loss.

But if we work instead from the power spectrum, we clearly see that the standard expression is obtained,
\begin{equation}
 \Delta^2_{\zeta}=\left[\frac{H^2}{\dot{\phi}^2}\frac{H^2}{(2\pi)^2}\right]_{k=aH}.
\end{equation}
And using the expression for $\epsilon$ for standard high energy inflation, we can reexpress this as,
\begin{equation}
 \Delta^2_{\zeta}\sim\left[\frac{H^3}{\epsilon}\right]_{k=aH}.
\end{equation}
We thus see that the standard expression for the spectral index is recovered,
\begin{equation}
 n_s-1= -3\epsilon - \eta.
\end{equation}

\section{Conclusions}
We have studied slow-roll inflation on a brane where there is energy loss into the bulk due to graviton emission. We have built on previous work \cite{BraxDavis} and have discovered another novel result. 

First, as obtained in \cite{BraxDavis}, there are two different branches of inflationary dynamics; one a slightly modified form of the standard slow-roll dynamics, while the other very nonstandard. In this paper, we have explicitly derived the form of the energy loss term from inflaton-to-graviton scattering and have obtained the precise dynamics of the two branches. 

Further, we find that the curvature perturbation on superhorizon scales is \emph{not conserved} even for the purely adiabatic perturbations generated in single-field inflation. While it is common for the curvature perturbation to not be conserved in multifield inflationary models due to the presence of isocurvature modes, we believe this is the first time such a phenomenon has been demonstrated for a model of single-field inflation. Indeed, nonconservation of curvature perturbation on superhorizon scales should be a general feature of any single-field inflation model whenever the (brane) energy-momentum tensor is not conserved.

The standard method of matching the late time power spectrum to the spectrum of quantum fluctuations at Hubble exit therefore no longer holds. The evolution of perturbations on superhorizon scales must be tracked until horizon reentry. To do this, we have adapted methods used in multifield inflationary models to our model of single-field inflation in a braneworld scenario. The power spectra and spectral indices obtained for the modified slow-roll branch and the nonstandard branch are quite different from the standard expressions for high energy inflation. But we note that the standard results for high energy inflation in braneworlds is recovered for the modified slow-roll branch in the limit $\alpha\rightarrow 0$.

\section*{Acknowledgments}
 It is a pleasure to thank Daniel Baumann, Philippe Brax, Carsten van de Bruck, Anne-Christine Davis, Baojiu Li, Eugene Lim, and Paul Shellard for valuable comments and discussions. We further thank Philippe Brax and Anne-Christine Davis for sharing some of their early work and for suggesting this problem. Ashok is supported by the Cambridge Commonwealth Trust, DAMTP, and Trinity College, Cambridge.

\end{document}